\documentclass{aa}

\newcommand{\av}{$A_V$}

\newcommand{\lsun}{L$_{\odot}$}
\newcommand{\msun}{M$_{\odot}$}
\newcommand{\rsun}{R$_{\odot}$}
\newcommand{\msunyr}{M$_{\odot}$\,yr$^{-1}$}
\newcommand{\macc}{$\dot{M}_{acc}$}

\newcommand{\lacc}{$L_{\mathrm{acc}}$}
\newcommand{\rstar}{$R_{\mathrm{*}}$}

\newcommand{\mstar}{$M_{\mathrm{*}}$}

\newcommand{\hi}{\ion{H}{i}}
\newcommand{\nai}{\ion{Na}{i}}
\newcommand{\cai}{\ion{Ca}{i}}
\newcommand{\hei}{\ion{He}{i}}
\newcommand{\feii}{\ion{Fe}{ii}}
\newcommand{\mgi}{\ion{Mg}{i}}
\newcommand{\fei}{\ion{Fe}{i}}
\newcommand{\nii}{\ion{Ni}{i}}
\newcommand{\oi}{\ion{O}{i}}
\newcommand{\azii}{\ion{N}{ii}}
\newcommand{\sii}{\ion{S}{ii}}
\newcommand{\caii}{\ion{Ca}{ii}}

\usepackage{graphicx}
\usepackage{txfonts}

\begin{document}

  \title{The 2016-2017 peak luminosity of the pre-main sequence variable V2492 Cyg}


   \author{T. Giannini\inst{1}, U. Munari\inst{2}, S. Antoniucci\inst{1}, D. Lorenzetti\inst{1}, A.A. Arkharov\inst{3}, S. Dallaporta\inst{4}, A. Rossi\inst{5}, G. Traven\inst{6}
 }

   \institute{INAF-Osservatorio Astronomico di Roma, via Frascati 33, I-00040 Monte Porzio Catone, Italy 
              \email{teresa.giannini@oa-roma.inaf.it}         
   \and           
   INAF - Osservatorio Astronomico di Padova, via dell' Osservatorio 8, 36012, Asiago (VI), Italy
   \and
   Central Astronomical Observatory of Pulkovo, Pulkovskoe shosse 65, 196140, St.Petersburg, Russia
   \and
   ANS Collaboration, Astronomical Observatory, 36012, Asiago (VI), Italy   \and 
   INAF - Osservatorio Astronomico di Bologna, via Ranzani 1, 40127, Bologna, Italy
   \and 
   Faculty of Mathematics and Physics, University of Ljubljana, Jadranska 19, 1000 Ljubljana, Slovenia     }
     
\date{Received; accepted }

  \abstract
{V2492 Cyg is a young pre-main sequence star presenting repetitive brightness variations of significant amplitude ($\Delta R$ $\ge$ 5 mag) whose physical origin has been ascribed to both extinction (UXor-type) and accretion (EXor-type) variability,  although their mutual proportion has not been clarified yet. 
Recently, V2492 Cyg has reached a level of brightness ever registered in the period of its documented activity.} 
{We aim to derive the variation of the mass accretion rate between low- and high-state and to get new insights on the origin of the variability of V2492 Cyg.}
{Optical and near-infrared photometry and spectroscopy have been obtained in October 2016 and between March and July 2017. The source has remained bright until the end of May 2017, then it started to rapidly fade since the beginning of June at a rate of $\sim$ 0.08 mag/day. On mid-July 2017 the source has reached the same low-brightness level as two years before. Extinction and mass accretion rate were derived by means of the luminosity of the brightest lines, in particular H$\alpha$ and H$\beta$. A couple of optical high-resolution spectra are also presented to derive information on the gas kinematics. }
{Visual extinction variations do not exceed a few magnitudes, while the mass accretion rate is estimated to vary from less than 10$^{-8}$ up to a few 10$^{-7}$ \msunyr. This latter is comparable to that estimated on the previous high-state in 2010, likely occurred under more severe extinction conditions.}
{The combined analysis of the optical and near-infrared (NIR) observations extends to the present event the original suggestion that the V2492 Cyg variability is a combination of changing  extinction and accretion.}

\authorrunning{T. Giannini et al.}
\titlerunning{The 2016-2017 peak luminosity of the pre-main sequence variable V2492 Cyg}

\maketitle
%

\section{Introduction}\label{sec:sec1}
Intermittent outbursts of significant amplitude ($\Delta$V $\sim$ 3-5 mag), due to repetitive accretion events onto pre-main sequence (PMS) stars, are referred as EXor phenomena (Herbig 1989, 2008; Hartmann, Herczeg, \& Calvet 2016). These bursts are thought to be triggered by instabilities in the inner
 parts of the circumstellar disk (D'Angelo \& Spruit 2010, 2012; Armitage 2016). Observationally, EXor are quite rare systems, with only a few tens of examples known so
  far (see e.g., Audard et al. 2014) and in some cases the classification is uncertain since the observed properties (burst amplitude, cadence, optical/IR colors) are attributable to accretion as well as to extinction episodes related to orbiting dust structures that move along the line of sight  (UXors variables, Grinin 1988). These latter are more compatible with repeating brightness variations with the same period and amplitude, while more erratic variations favor the accretion hypothesis. This is the reason why a photometric and spectroscopic monitoring of subsequent outbursts are in order.

In the framework of our monitoring program EXORCISM (Antoniucci et al. 2014), an ideally suited candidate is the young low-mass object V2492 Cyg located in the Pelican Nebula ($\alpha_{2000}$ = 20$^{h}$51$^{m}$26.23$^{s}$, $\delta_{2000}$ =  +44$^{\circ}$05$^{\prime}$23.9$^{\prime \prime}$), which has shown  repetitive brightness variations with $\Delta R$ $\ga$ 5 mag. The first recorded episode was discovered  on 2010 August 23 by Itagaki \& Yamaoka (2010).
This event has been
followed by brightening and fading events carefully investigated by Hillenbrand et al. (2013), and interpreted as a quasi-periodical phenomenon consistent with changes of extinction in the inner 
disk material ($\Delta$\av $\la$ 30 mag). The role of dust obscuration  was further enhanced by the study of K\'{o}sp\'{a}l et al. (2013). Based on the 
constancy of the far-infrared fluxes, these authors conclude that the observed variability is due to the occultation of the inner disk by a dense, long-live dust cloud, and therefore they regarded V2492 Cyg as a young UXor variable. From a spectroscopic point of view, however, V2492 Cyg  presents features more typical of EXor spectra, such as \hi\, lines and CO overtone bandheads in emission, as well as signatures of a well developed outflow (Covey et al. 2011, Aspin 2011, Hillenbrand et al. 2013). Therefore, a unified picture of the physical phenomena occuring on V2492 Cyg and in its environment is still lacking.

The most recent  (November 2016-March 2017) peak of brightness, the highest ever recorded, has been announced by Ibryamov \& Semkov (2017) and  photometrically monitored by Froebrich et al. (2017) and Munari et al. (2017).
Here we present  optical and near-infrared (NIR) photometry and spectroscopy taken during this phase of maximum brightness along with the subsequent fading currently ongoing. Observations are described in Section 2, while the results are presented and discussed in Section 3. Our summary is given in Section 4.

\section{Observations}\label{sec:sec2}
\subsection{Photometry}\label{sec:sec2.1}
\begin{figure*}
   \centering
   \includegraphics[width=14cm]{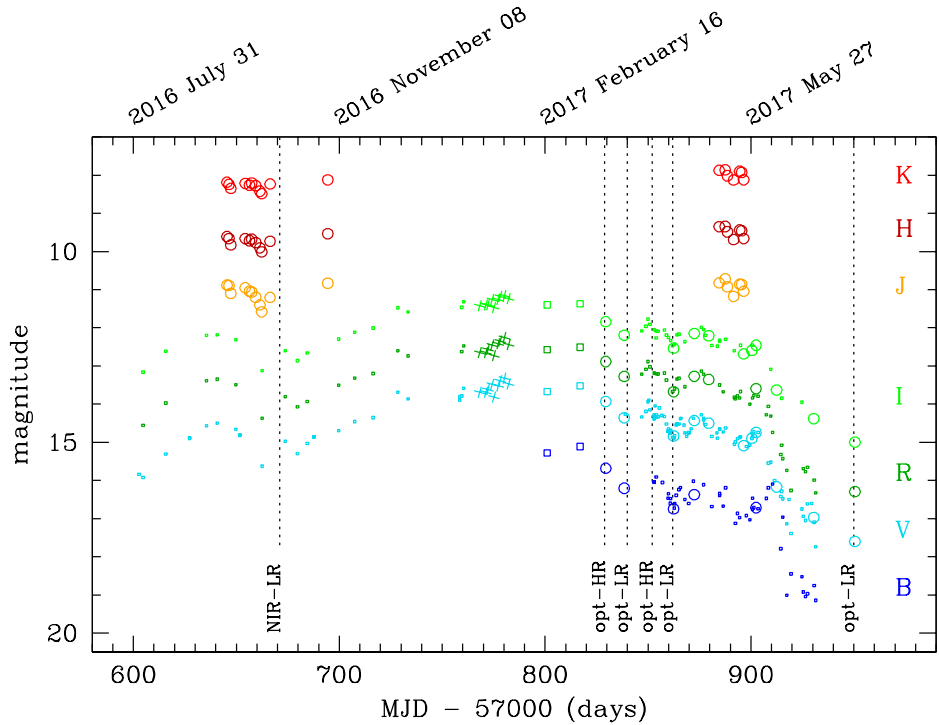}
   \caption{Optical and NIR light curve. Our new photometric data are depicted as large  open circles, while other points are from AAVSO (dots); Froebrich et al. 2017 (crosses), and Ibryamov \& Semkov 2017 (open squares). Vertical dotted lines indicate the dates of the spectroscopic observations. \label{fig:fig1}}
   \end{figure*}

$BVR_{\rm C}I_{\rm C}$ optical photometry (Table\,\ref{tab:tab1} and Figure\,\ref{fig:fig1}) has been
obtained with the telescope ID 310 operated in Cembra (Italy) by the Asiago Novae and Symbiotic stars (ANS) Collaboration.
Data reduction involved correction for nightly bias, dark and flat fields. Full photometric calibration was achieved from APASS survey data 
(Henden et al.  2012, Henden \& Munari 2014) using the SLOAN-Landolt transformation equations
calibrated in Munari (2012) and Munari et al.  (2014a,b).

$JHK$ photometry (Table\,\ref{tab:tab2} and Figure\,\ref{fig:fig1}) was obtained at the 1.1m AZT-24 telescope located at
Campo Imperatore (L'Aquila - Italy) equipped with the
imager and spectrometer SWIRCAM (D'Alessio et al. 2000). All the observations were obtained by dithering
the telescope around the target position. The raw imaging data
were reduced by using standard procedures for bad pixel removal, flat fielding, and sky subtraction. Photometric calibration was
obtained from the 2MASS photometry of bright stars present in the  4$\farcm$4 $\times$4$\farcm$4 field.

%

\begin{table}
\caption{\label{tab:tab1}Optical photometry.}
\centering
\begin{tabular}{cccccc}
\hline\hline
MJD        &    Date    & $B$      &   $V$    &$R_{\rm C}$&$I_{\rm C}$ \\\hline
57829      & 2017 03 17 &  15.68  &  13.93  &   12.88  &   11.84   \\
57838      & 2017 03 26 &  16.20  &  14.36  &   13.27  &   12.19   \\ 
57862      & 2017 04 19 &  16.75  &  14.83  &   13.67  &   12.53   \\
57872      & 2017 04 29 &  16.37  &  14.43  &   13.27  &   12.15   \\
57879      & 2017 05 05 &    -    &  14.50  &   13.36  &   12.21   \\
57896      & 2017 05 23 &    -    &  15.09  &    -     &   12.68   \\
57900      & 2017 05 27 &    -    &  14.89  &    -     &   12.59   \\
57902      & 2017 05 29 &  16.71  &  14.72  &   13.59  &   12.46   \\
57912      & 2017 06 07 &    -    &  16.17  &    -     &   13.63   \\
57930      & 2017 06 25 &    -    &  16.97  &    -     &   14.38   \\
57950      & 2017 07 16 &    -    &  17.60  &   16.29  &   15.00   \\
 \hline
\end{tabular}
\tablefoot{Errors are about 0.02 mag.}
\end{table}
  
\begin{table}
\caption{\label{tab:tab2}NIR photometry.}
\centering
\begin{tabular}{ccccccccc}
\hline\hline
MJD        &    Date    &    J &   H    &   K\\
\hline
57645      & 2016 09 14 &  10.88    &  9.61      &  8.18   \\  
57646      & 2016 09 15 &  10.89    &  9.66      &  8.24   \\
57647      & 2016 09 16 &  11.10    &  9.83      &  8.35   \\
57654      & 2016 09 23 &  10.95    &  9.66      &  8.21   \\
57656      & 2016 09 25 &  11.04    &  9.71      &  8.26   \\
57657      & 2016 09 26 &  11.06    &  9.68      &  8.21   \\
57659      & 2016 09 28 &  11.20    &  9.76      &  8.28   \\
57661      & 2016 09 30 &  11.40    &  9.90      &  8.42   \\
57662      & 2016 10 01 &  11.58    & 10.01      &  8.48   \\
57666      & 2016 10 05 &  11.20    &  9.73      &  8.23   \\
57694      & 2016 11 02 &  10.83    &  9.53      &  8.12   \\
57884      & 2017 05 11 &  10.82    &  9.35      &  7.87   \\
57887      & 2017 05 14 &  10.72    &  9.34      &  7.86   \\
57888      & 2017 05 15 &  10.92    &  9.48      &  8.01   \\
57891      & 2017 05 18 &  11.17    &  9.44      &  8.13   \\
57894      & 2017 05 21 &  10.86    &  9.46      &  7.90   \\
57895      & 2017 05 22 &  10.86    &  9.46      &  7.93   \\
57896      & 2017 05 23 &  11.04    &  9.66      &  8.12   \\
\hline
\end{tabular}
\tablefoot{Errors are less than 0.03 mag.}
\end{table}

\begin{table}
\caption{\label{tab:tab3} Journal of the spectroscopic observations.}
\centering
\begin{tabular}{ccccc}
\hline\hline            
MJD        &    Date    &  Instr/Tel  & $\Re$ & T$_{int}$\\
           &(yyyy mm dd)&             &       & (s)  \\
\hline
57197      & 2015 06 24 &   MODS/LBT  & 1500    & 1500    \\
57665      & 2016 10 04 &   LUCI2/LBT & 1000    & 1200    \\  
57829      & 2017 03 17 &   ECH/1.82m &18\,000  & 2400    \\
57840      & 2017 03 28 &  B\&C/1.22m & 2400    & 600     \\
57852      & 2017 04 09 &   ECH/1.82m & 18\,000 & 4200    \\
57862      & 2017 04 19 &  B\&C/1.22m & 2400    & 900     \\
57950      & 2017 07 16 &  B\&C/1.22m & 2400    & 1800    \\
\hline
\end{tabular}
\end{table}

\subsection{Spectroscopy}\label{sec:sec2.2}
\begin{figure*}
   \centering
   \includegraphics[width=14cm]{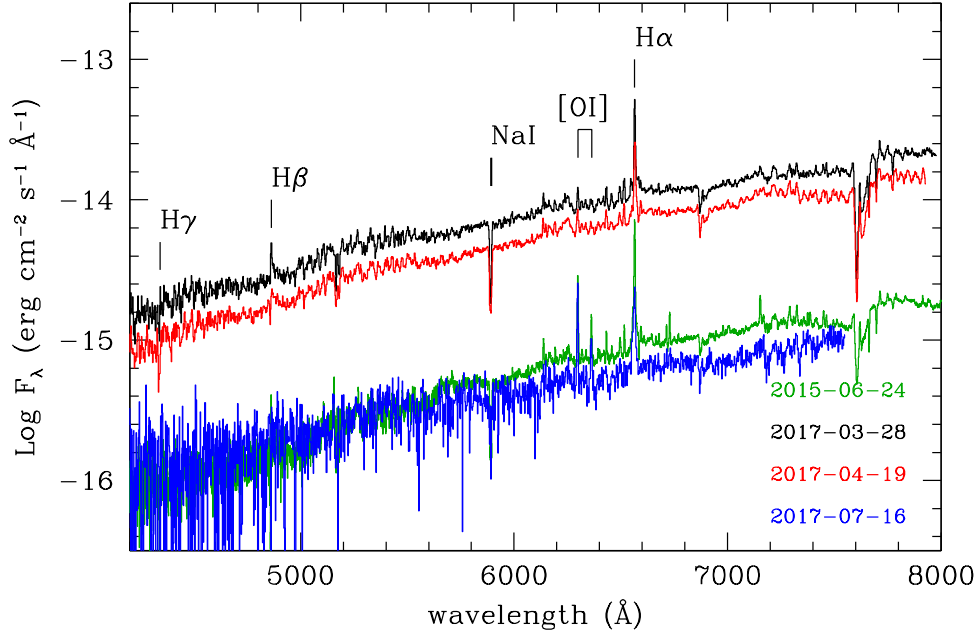}
   \caption{Optical low-resolution spectra of V2492 Cyg in the four dates reported on the bottom-right. Black, red and blue\,: 1.22m/B\&C; green: LBT/MODS. Main emission/absorption lines are labeled.\label{fig:fig2}}
    \end{figure*}

Table\,\ref{tab:tab3} gives the journal of the spectroscopic observations. We obtained four optical low-resolution spectra of V2492 Cyg, shown in 
Figure\,\ref{fig:fig2}. The first one was taken using the Multi-Object Double Spectrograph (MODS - Pogge et al. 2010) mounted at  the Large Binocular Telescope (LBT) on 2015 June 24, when the source  was at a low level of activity
(AAVSO\footnote{{\it American Association of Variable Star Observers}, (https://www.aavso.org/)} magnitudes of 2015 June 22: $V$ =17.6, $R$=16.0, $I$=14.5). We integrated 1500 s in the spectral range 3200$-$9500~\AA\, by using a 0$\farcs$8 slit ($\Re \sim$ 1500).
Other three low-resolution spectra were obtained on 2017 March 28,  2017 April 19, and 2017 July 16, being the first two very close in time to the brightness peak, and the last one obtained when the source was already faded. These spectra 
($\sim$\,3500$-$8000~\AA\, at $\Re\sim$\,2400) were obtained with the 1.22\,m telescope + B\&C spectrograph (Asiago, Italy) operated by the University of Padova.
The steps adopted for of the data reduction are: correction for dark and bias, bad-pixel mapping, flat-fielding, and extraction of one-dimensional spectrum by integrating the stellar trace along the spatial direction. Wavelength calibration was obtained from the spectra of arc lamps. The Asiago spectra have been calibrated in flux through observations of standards before and after V2492 Cyg, while the MODS spectrum was calibrated by using the AAVSO magnitudes given above.
 
\begin{figure*}
   \centering
   \includegraphics[width=14cm]{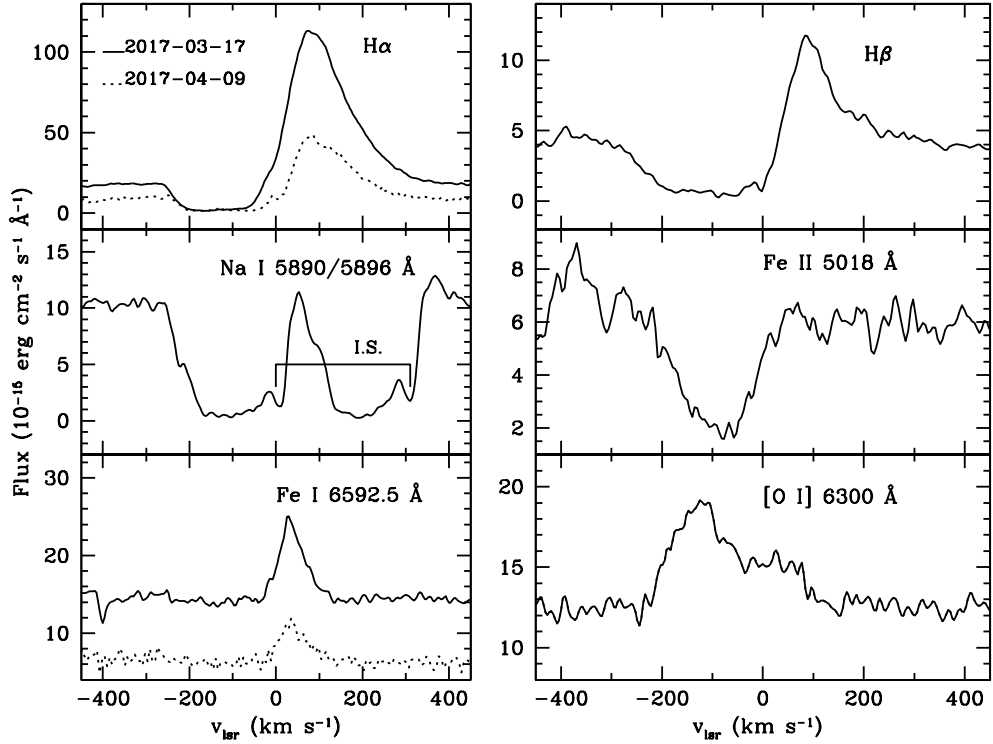}
   \caption{Examples of line profiles in the Echelle spectrum obtained at the reported dates. Interstellar absorption features are marked in the \nai\, profile. \label{fig:fig3}}
    \end{figure*}

\begin{figure*}
   \centering
  \includegraphics[width=14cm]{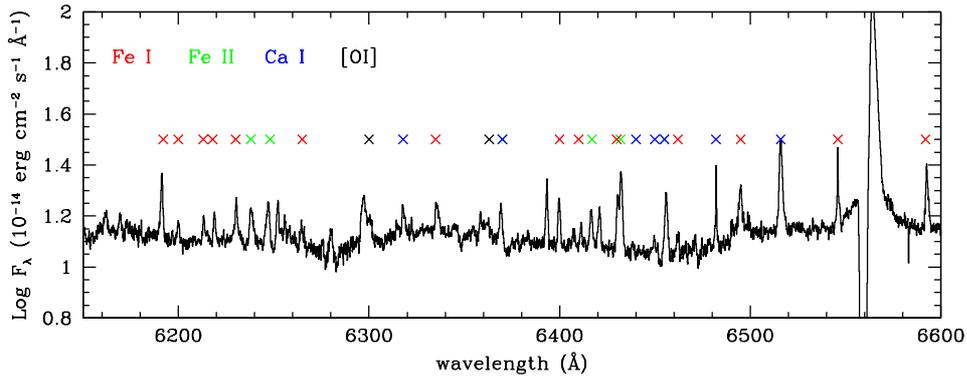}
   \caption{A sample segment from our Echelle spectrum of March 17, 2017 that
 highlights the rich, low ionization emission spectrum presented by V2492 Cyg.\label{fig:fig4}}
    \end{figure*}

Two high-resolution optical spectra at  $\Re$\,$\sim$\,18\,000 were acquired on 2017 March 17 and 2017 April 9 in the range 3300$-$8050~\AA\, with the REOSC Echelle spectrograph mounted on the Asiago 1.82\,m telescope. Data reduction and calibration were performed following 
the same steps cited above. Profiles of  interesting lines are shown in Figure\,\ref{fig:fig3} while a sample interval of the spectrum of March 17 is depicted in Figure\,\ref{fig:fig4}.

\begin{figure*}
   \centering
   \includegraphics[width=14cm]{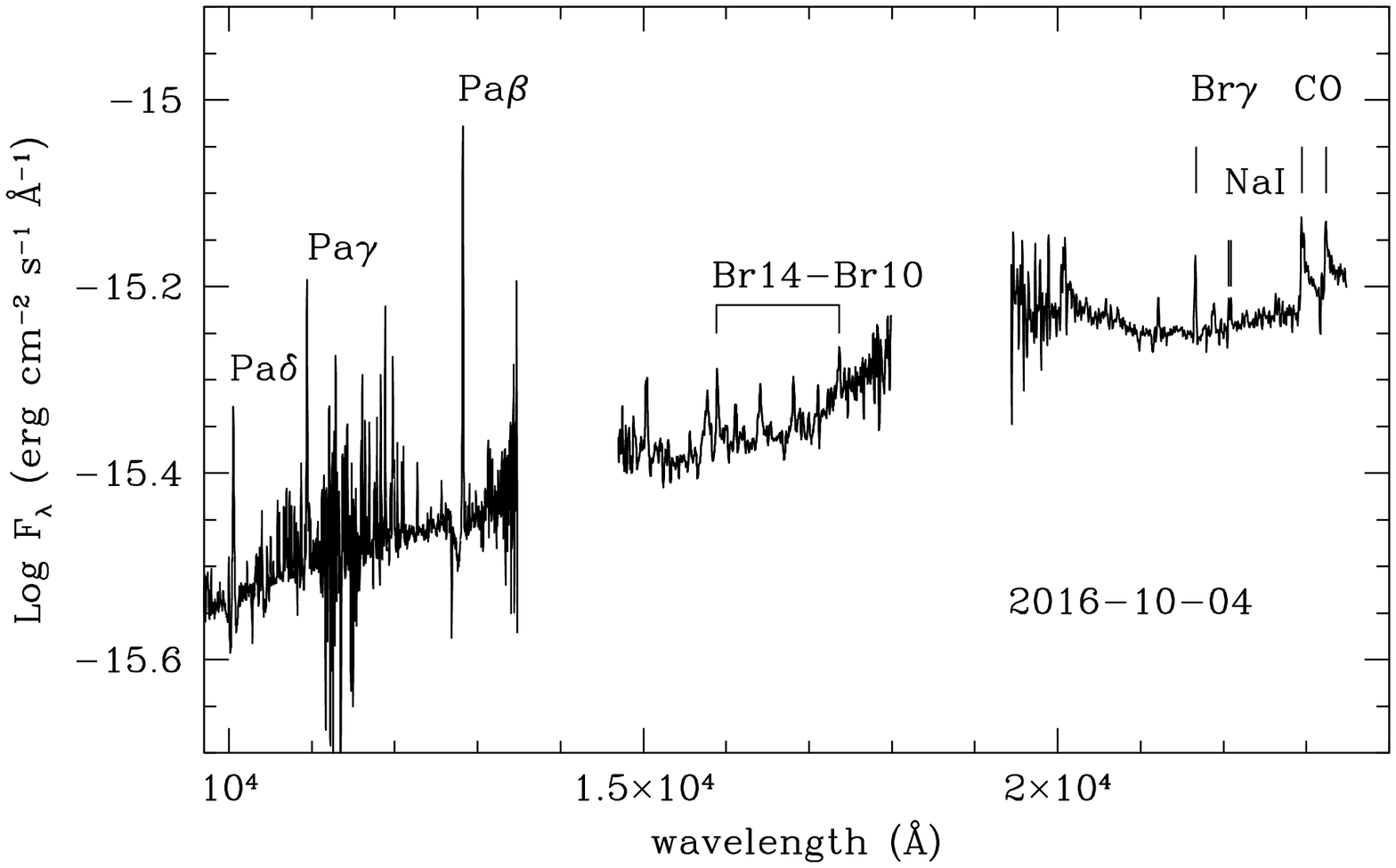}
   \caption{Near-infrared LBT/LUCI2 spectrum. Main emission lines are labeled.\label{fig:fig5}}
    \end{figure*}

A NIR spectrum at $\Re$ $\sim$\,1300  was obtained with LUCI2  at LBT on 2016  October 4, when the source was already in a high-state of activity (Figure\,\ref{fig:fig5}). The observations were carried out with the G200 low-resolution grating coupled with the 0$\farcs$75 slit. The standard ABB'A' technique was adopted to perform the observations using the $zJ$ and $HK$ grisms (1.0$-$2.4\,$\mu$m), for total integration times of 12 and 8 minutes, respectively.
The raw spectral images were flat-fielded, sky-subtracted, and corrected for optical distortions in both the spatial and spectral directions. Atmospheric absorptions were removed using the normalized spectrum of a telluric standard star, after fitting its intrinsic spectral features. Wavelength calibration was obtained from arc lamps, while for flux calibration the photometry of the day after the spectroscopic observation was assumed. No intercalibration was performed between the zJ and HK parts of the spectrum, since they were already optimally aligned.

%
%

\section{Results and discussion}\label{sec:sec3}
\subsection{Light curve}\label{sec:sec3.1}

Since its first recorded outburst in 2010, V2492 Cyg has continuously shown a high degree of photometric variability (see e.g., the AAVSO light-curve). In the present paper we concentrate on the last maximum phase and subsequent declining from September 2016  to July 2017 (MJD from 57645 to 57950).
In  Figure~\ref{fig:fig1}, together with our photometric data, we show also some literature (Ibryamov \& Semkov 2017; Froebrich et al. 2017) and AAVSO data that optimally trace the brightness evolution. 
The peak of this last event corresponds to $V$ $\simeq$ 13.5 mag. Previous brightness enhancements, monitored during the last seven years, peaked between 16 and 14 mag in the $V$ band (AAVSO light-curve).  In the "extinction-driven" scenario, the light-curve V2492 Cyg roughly resembles that of long-term  (years to decades), quasi-cyclic UXor variables, such as, for example, BF Ori (Herbst \& Shevchenko 1999), albeit in these sources the amplitude difference between subsequent peaks is definitely lesser (typically a few tens of magnitude) than in V2492 Cyg.

Our NIR monitoring does not cover exactly the five months of maximum brightness, but it significantly samples the peak activity period (September 2016 - May
 2017, MJD from 57645 to 57896) during  which the average magnitude fluctuations did not exceed 0.2 mag in all the NIR bands. An exception is the local
  brightness drop occurred between MJD=57659 and MJD=57666 (peak minimum at MJD=57662), registered also in the optical AAVSO light-curve. After this event, the NIR brightness has returned to the previous level and then remained fairly stable until the last photometric observation acquired in May 2017.

The optical monitoring started on March 2017 (MJD=57829), immediately after the brightness peak, and extended  during the course of the declining phase, roughly started on April 2017 and still ongoing at the date of the last photometric point on 16 July 2017 (MJD=57950). From the beginning of April to the end of May 2017 the 
source has slowly faded at a rate of $\sim$ 0.01 mag/day, then, since the beginning of June, it started a quick declining at $\sim$\,0.08 mag/day, as also shown by recent AAVSO data. This latter speed is similar to that of $\sim$\,0.1 mag/day measured by Hillenbrand et al. (2013) in three larger brightness drops occurred between 2010 and 2012. \\

\subsection{Photometric diagrams}\label{sec:sec3.2}

\begin{figure*}
   \centering
   \includegraphics[width=17cm]{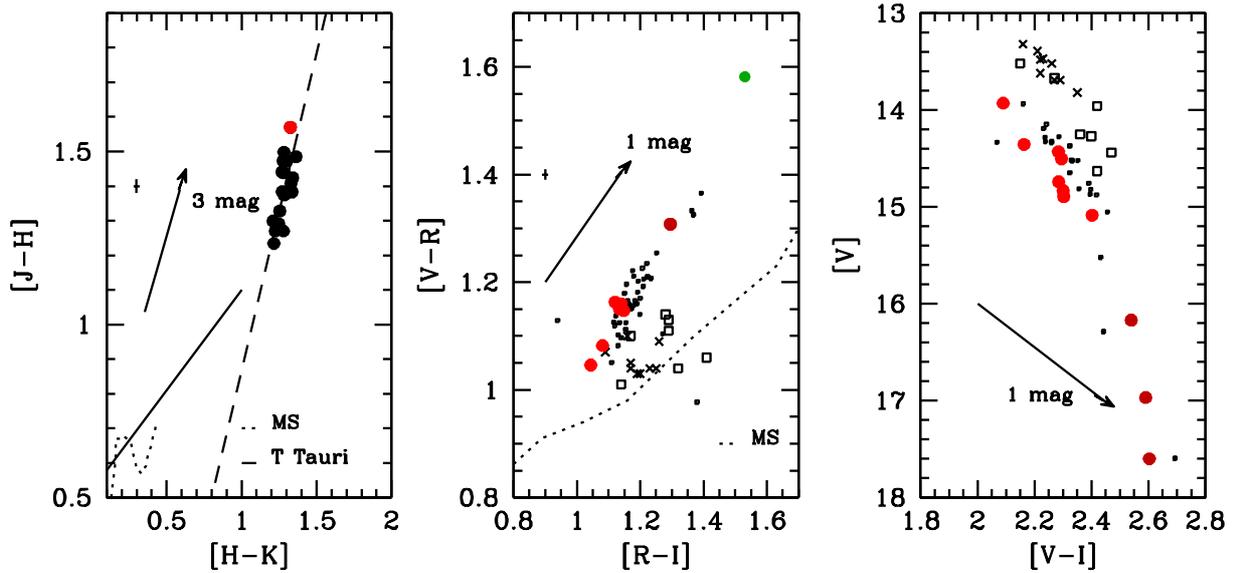}
   \caption{Left panel: NIR two-color plot [$J-H$] vs. [$H-K$]. The red point represents the NIR colors at MJD=57662 (see text). Continuous line: locus of T Tauri stars (Meyer et al. 1997). Dotted line: unreddened main-sequence. Dashed line: linear fit through the data. Arrow: extinction vector corresponding to A$_V$ = 3 mag (reddening law of Cardelli et al.1989). The average error is indicated with a cross. Middle panel: optical two-color plot [$V-R$] vs. [$R-I$].  Arrow: extinction vector corresponding to A$_V$ = 1 mag. Dotted line: unreddened main-sequence. Red and dark-red large dots: our data taken before and after June 2017. The green dot represents the colors of 2015 June 22 (AAVSO database). Black data : literature/AAVSO data with the same symbols as in Figure\,\ref{fig:fig1}. Right panel: [$V$] vs. [$V-I$]. Symbols and colors as in the middle panel.\label{fig:fig6}}
   \end{figure*}

In Figure~\ref{fig:fig6} the photometric data are presented in form of two-colors and magnitude-color plots: 
$[J-H]$ vs. $[H-K]$ (left panel),  $[V-R]$ vs. $[R-I]$ (middle panel), and [$V$] vs. [$V-I$] (right panel). 
NIR colors align well with the extinction vector (law of Cardelli et al. 1989). Their spread corresponds to a $\Delta A_V$ between 1 and 3 mag, being the reddest colors those of the local brightness minimum  occurred at MJD=57662 (see Figure\,\ref{fig:fig6}).
This confirms the overall picture derived by the NIR color variations between 2010 and 2012 over a much larger range of magnitudes ($\Delta J$ $\sim$ 5 mag), corresponding to $\Delta A_V$  $\sim$ 30 mag (Hillenbrand et al. 2013).

The optical colors, however, do not fully agree with extinction variability alone. While during the rising and declining phases color variations are consistent with A$_V$ changes $\la$\,1 mag,  at the brightness peak $[R-I]$ becomes significantly redder of about 0.4 mag, while $[V-R]$ maintains roughly constant (see the data of Froebrich et al. 2017 and Ibryamov \& Semkov 2017 in Figure\,\ref{fig:fig6}). This behavior is similar to that exhibited by the optical colors of the EXor source V1118 Ori during its  2015 outburst (Figure 7 of Giannini et al. 2017).  

An even more evident deviation from the reddening vector is evident in the [$V$] vs. [$V-I$] data points taken between June and July 2017 (right panel of Figure\,\ref{fig:fig6}\,).
This changing in slope is likely contributed by both extinction and accretion rate variations, whose combined effect might also be responsible for the acceleration in the declining speed from 0.01 to 0.08 mag/day (Sect.\ref{sec:sec3.1}), observed in the same period. Noticeably, the observed feature is characteristic of a number of objects, like V1180 Cas (Kun et al. 2011), V1647 Ori (Mosoni et al. 2013), and in particular V346 Nor, where accretion rate and extinction enhancements appear significantly correlated (K{\'o}sp{\'a}l et al. 2017). Also, the photometric variability of the pre-main sequence source V1184 Tau has been interpreted by Grinin et al. (2009) as due to repetitive accretion bursts followed by an increase of the geometric thickness  of the  inner disk rim, which is able to obscure a relevant portion of the central star.

\begin{table*}
\caption{\label{tab:tab4} Fluxes of diagnostic lines.}
\centering
\begin{tabular}{cccccccc}
\hline\hline            
 Date    &     \multicolumn{6}{c}{F$\pm\Delta$F }                                                                               \\
(yyyy mm dd)&  \multicolumn{6}{c}{(10$^{-14}$ erg s$^{-1}$ cm$^{.2}$)}                                                          \\
\hline
           &    H$\alpha$          &  H$\beta$                &  Pa10        & Pa9          & \hei\,6678\AA   & [\oi]6300\AA    \\
2015 06 24 &   5.07$\pm$0.04       &  0.06 (-0.03)$\pm$0.01   &  0.65$\pm$0.2& 0.7$\pm$0.2  & 0.1 $\pm$0.03   & 1.25$\pm$0.04   \\
\hline           
           &  Pa$\delta$          &   Pa$\gamma$              &  Pa$\beta$   & Br$\gamma$    &                &                 \\
2016 10 04 &  11.8 (-1.2)$\pm$0.3  &  20.1$\pm$0.4            & 29.1$\pm$0.4 &12.1$\pm$0.4   &                &                 \\  
\hline
           &  H$\alpha$           &   H$\beta$                &  H$\gamma$    &  \caii\,-K     & \hei\,6678\AA  &  [\oi]6300\AA \\ 
2017 03 17 & 41.7 (-5.0)$\pm$0.2  &   1.9 (-1.1)$\pm$0.2      & -0.7$\pm$0.2  &      -       &  0.8 $\pm$0.1  & 2.7$\pm$0.2     \\
2017 03 28 & 39.8$\pm$0.3         &   2.6$\pm$0.2             &    -          & 0.9$\pm$0.3  &  1.0 $\pm$0.2  & 2.2$\pm$0.4     \\
2017 04 09 & 15.4 (-2.2)$\pm$0.3  &   -                       &    -          &              &  0.5 $\pm$0.1  &  -              \\
2017 04 19 & 23.4$\pm$0.3         &   1.4$\pm$0.3             &    -          &  $<$1        &    -           & 2.0$\pm$0.3     \\
2017 07 16 & 2.1$\pm$0.1          &   -                       &    -          &   -          &    -           & 1.3$\pm$0.1     \\
\hline
\end{tabular}
\end{table*}

\subsection{Description of the observed spectra}\label{sec:sec3.3}

The 2010$-$2012 optical-NIR spectrum of V2492 Cyg has been studied in detail by  Aspin (2011) and Covey et al. (2011), both in low- and high-brightness phases. It
 was rich of forbidden, permitted, and recombination atomic lines as well as molecular lines.  In our low-resolution optical spectra  of March and April 2017 the  brightest lines  are H$\alpha$, H$\beta$, H$\gamma$, and \nai\,5890/6 \AA\,, with H$\gamma$ and \nai\, seen in absorption (Figure\,\ref{fig:fig2}) . In these spectra [\oi]6300 \AA\, is also identified at signal-to-noise ratio of about five, while, together with H$\alpha$, it becomes the brightest line in the quiescence spectra of  
  June 2015 and July 2017. In addition, some \hi\, lines of the Paschen series and \hei\,6678\AA\, are detected in the MODS spectrum of June 2015.  Permitted lines emission of neutral and singly ionized species (mainly \fei\,, \feii\,, and \cai\,) is also detected in our high-resolution optical spectra of March and April 2017 (see Figure\,\ref{fig:fig4}). 

In the NIR the most prominent lines are the \hi\, lines of the Paschen and Brackett series (Figure\,\ref{fig:fig5}). We also report the detection of the  \nai\,2.206, 2.207 $\mu$m doublet along with CO v=2-0 and v=3-1 bandhead emission. Noticeably, in EXor  spectra \nai\, and CO are seen both in absorption or emission depending on the level of the source activity. Examples are, among others, XZ Tau, PV Cep, V1647 Ori, V1118 Ori  (Lorenzetti et al. 2009, Aspin et al. 2008, Giannini et al. 2017), and V2492 Cyg itself (Hillenbrand et al. 2013).

In Table\,\ref{tab:tab4} we give the fluxes of the lines that we have used for the subsequent analysis, and in particular for the determination of the mass accretion rate.

\subsection{Mass accretion rate determination}\label{sec:sec3.4}

\begin{table}
\caption{\label{tab:tab5} Extinction, accretion luminosity and mass accretion rate.}
\begin{tabular}{cccc}
\hline\hline            
 Date    &     \av   &  \lacc$\pm\Delta$\lacc      & \macc$\pm\Delta$\macc\\
(yyyy mm dd)&  (mag) &   (\lsun)                   &   (10$^{-7}$ \msunyr)  \\
\hline
2015 06 24 &   4     &  0.4$\pm$0.1                &    0.65$\pm$0.15  \\2016 10 04 &  2.5    &  4.4$\pm$1,1                &    5.3$\pm$0.9    \\  
2017 03 17 &   3     &  1.4$\pm$0.7                &    1.6$\pm$0.8    \\2017 03 28 &  3.5    &  2.8$\pm$0.5                &    2.2$\pm$0.8    \\2017 04 09 &   3     &  0.9$\pm$0.3                &    0.6$\pm$0.3    \\2017 04 19 &   3     &  0.7$\pm$0.2                &    0.8$\pm$0.3    \\2017 07 16 &2.5$-$5\tablefootmark{a}&  0.02$-$0.3  &    0.03$-$0.3     \\\hline
\end{tabular}
\tablefoottext{a}{Assumed range.}
\end{table}

The computation of the mass accretion rate \macc\ is based on the empirical relationships by Alcal\'a et al. (2017), which connect the accretion luminosity \lacc\, to the luminosity of many emission lines ($L_{line}$) in the optical and near-IR range. These relationships were obtained from VLT/X-Shooter observations of a large sample of Class II sources in Lupus, by modeling the emission observed in excess to the stellar photosphere as the continuum emission of a slab of hydrogen, which provides a independent measurement of \lacc\ (see Alcal\'a et al. 2014, Manara et al. 2013).
Alcal\'a et al. (2017) provide relations for all the \ion{H}{i} lines observed in our V2492 Cyg spectra, plus the \ion{He}{i} 6678\AA\, and Ca II K lines.

From each of these lines we are able to derive a value of \lacc, provided that the distance and extinction are known, as we need to convert the measured line fluxes into luminosities. For the distance to the Pelican Nebula, we considered the value 550 pc (Bally \& Reipurth 2003). Since V2492 Cyg is subject to a highly
variable extinction, we preferred not to assume a previously determined estimate of \av\,, so we adopted the following method.
We considered a range of values for \av\ (from 0 to 30 mag in steps of 0.5 mag) and for each of these values we computed \lacc\ from each line by using the relevant relationship by Alcal\'a et al. (2017). This way, we obtained different sets of accretion luminosities (one for each \av\ value) that are characterized by different dispersions. The extinction for which the obtained \lacc\ dispersion shows a minimum is the \av\ value we eventually adopted.
Once the extinction was thus determined, we computed \lacc\ as the median value of the accretion luminosities of this least dispersed set (see e.g., Giannini et al. 2016, 2017).
The \av\, and \lacc\ values derived from this procedure are reported in Table\,\ref{tab:tab5}. In particular, we derived \av=4 mag in June 2015, and between 2.5$-$3.5 mag in March and April 2017, which is in reasonable agreement with the indications of the color-color diagrams displayed in Figure\,\ref{fig:fig6}.
The previous method could not be applied to estimate \av\ on the spectrum of July 2017, since only H$\alpha$ is detected there.
A reasonable range of \av\ is derived in this case by taking as lower limit the same value as in the previous dates (\av\ = 2.5 mag) and as upper limit
the one derived by assuming that the observed fade is caused only by the extinction increase (\av $\sim$ 5 mag, see Figure\,\ref{fig:fig1}).

Finally, we adopted the relation by Gullbring et al. (1998) to convert \lacc\, into \macc\,:
\begin{equation}
\label{eq:macc}
\dot{M}_{acc} = \frac{L_{acc}R_{*}}{GM_{*}} \left(1-\frac{R_{*}}{R_{in}}\right)^{-1} ~,
\end{equation}
where \mstar\ and \rstar\ are the stellar mass and radius
and $R_{in}$ is the inner truncation radius of the disk. We consider \rstar\ = 2\,\rsun\ and \mstar\ = 0.7\,\msun, as evaluated from
the EXor HR diagram recently published by Moody \& Stahler (2017), while for the factor $1-R_{*}/R_{in}$ we
assumed a value of 0.8, which implies a typical inner disk radius $R_{in}= 5 R_{*}$.

The derived \macc\, are reported in the final column of Table\,\ref{tab:tab5}. We find \macc\, $\sim$ 1.5$-$5 10$^{-7}$ \msunyr\, during the high-brightness phase,  with fluctuations that roughly follow the photometric evolution. The highest \macc\, determination is derived on October 2016. Since it was derived from NIR lines only, however, we cannot exclude that it might be biased, for example, by an underestimate of the \av\,, to which the NIR lines are less sensitive than the optical lines,  and/or an overestimate of \lacc\,. 

For comparison, we applied the method described above to the  \hi\, fluxes of two optical spectra obtained between August and October 2010 by 
Hillenbrand et al. (2013) and Aspin et al. (2011). We obtained A$_V$=5$-$9 mag and \macc\,$\sim$ 2 10$^{-7}$\,\msunyr\,, in optimal agreement with the estimates (A$_V \sim 6-$12 mag and \macc\,$\approx$ 2.5 10$^{-7}$\,\msunyr\,)\, given by Covey et al. (2011). Therefore, we can conclude that the 2010 and 2016  brightness enhancements have occurred at a similar level of the intrinsic luminosity but under different obscuration conditions.

As far as the low-brightness spectra is concerned, MODS lines observed in June 2015 are consistent with \macc\,$\sim$ 5$-$8 10$^{-8}$ \msunyr\,, roughly a factor of approximately ten or less lower than the peak value.  Conversely, a significant drop of \macc\, might have occurred on July 2017. A more precise A$_V$ estimate, however, should have been derived to make this result more reliable.

Summarizing, at variance for example with the case of V346 Nor, it seems that in V2492 Cyg there is not a simultaneous increase of  mass accretion rate and extinction. Rather, there are evidences for the opposite trend. We remark, however, that our data do not cover the entire variability range displayed by V2492 Cyg in the past (the minimum registered  $V$ is $\ga$ 20 mag, AAVSO light-curve). Observations during a deep minimum should be of great help to solve the question.

\subsection{Line profiles}\label{sec:sec3.5}
The two Echelle spectra, obtained at about one a month interval from each other, significantly differ, being the spectrum of March 17 much brighter. 
In this latter, the Balmer lines from H$\alpha$ to H$\delta$ are detected. H$\gamma$ and H$\delta$ are seen only in absorption while H$\alpha$ and H$\beta$ present composite profiles with a broad emission component slightly red-shifted and a flattened  P-Cyg  absorption extending up to $-$200 km s$^{-1}$. The spectrum is rich also in  low-lying ionized metallic lines (mainly \fei\,, \feii\,, and \nii\,, as depicted in Figure\,\ref{fig:fig4}). 
Their  FWHM ranges between 45 to 85 km s$^{-1}$ depending on ion and multiplet. This is typical of gas 
in funnel flows  close to the disk surface, where photons from the central star are partially shielded (Sicilia-Aguilar et al. 2012, K{\'o}sp{\'a}l et al. 2011, Beristain et al. 1998). All the lines are seen in emission, with the exception of \feii\, lines of multiplet 42 at 4923.93\AA\,, 5018.45\AA\,, 5169.03\AA\,, and possibly \mgi\, 5183.60\AA\, (Figure\,\ref{fig:fig3}). These lines present  absorption profiles centered at v$_{\rm lsr}$\,=\,$-$80 km s$^{-1}$, FWHM = 154 km s$^{-1}$ and a blue-end velocity similar to that of Balmer lines and \nai\, doublet. Their profiles are therefore suggestive of an accelerated and ionized gas outflowing in proximity of the star. Conversely to the case of V2492 Cyg, the \feii\, lines of multiplet 42 have been observed in emission on the spectra of the EXors DR Tau (Beristain et al. 1998) and EX Lup (Sicilia-Aguilar et al. 2012). In the latter source in addition to the main component in emission they also present a  blue-shifted absorption consistent with a wind origin. 

The [\oi]6300\AA\, line is also observed. It presents a double component, one roughly centered at the rest velocity with FWHM $\sim$ 144 km s$^{-1}$, and a blue-shifted component peaking at about $-$100 km s$^{-1}$ and extending up to $\sim$ $-$200  km s$^{-1}$, in agreement with the permitted wind lines. The  
[\oi]6363\AA\, is detected at a lower signal-to-noise ratio, while the [\azii] and [\sii] nebular emission lines reported, among others,
by Aspin (2011) and by Covey et al. (2011), are absent on our spectra.



\section{Summary}\label{sec:sec4}
We have presented optical and NIR photometric and spectroscopic observations of the pre-main sequence variable V2492 Cyg, obtained during the peak luminosity reached on 2016-2017. The main results of our observations can be summarized as follows:
\begin{itemize}
\item[-] One cause of the observed brightening is likely a strong decline of the local obscuration, with the estimated extinction definitively lower than during faint states (when A$_V$ may increase up to 30 mag). 
\item[-] A significant enhancement of the mass accretion rate has also likely occurred during the investigated period. This is testified by the deviation of the optical colors from the extinction law direction, the detection of \nai\, 2.2 $\mu$m and CO bands both in emission, and the enhancement of the \hi\ fluxes in the high-brightness spectra, and by the variation of the mass accretion rate obtained from the observed line luminosity. The existence of gas in funnel flows is also supported by the detection of metallic lines with width $\ga$ 50 km s$^{-1}$.  
\item[-] We obtain \macc\, of $\sim$ 2$-$5 10$^{-7}$ \msunyr\, at the peak, namely about a factor up to ten higher than in a fainter state prior the brightness enhancement. Conversely, our analysis suggests that a rapid drop of \macc\, is ongoing since July 2017.
The \macc\, value estimated at the brightness peak is similar to that reached in 2010, this latter likely occurred under more severe obscuration conditions.
\item[-] A high-velocity wind is probed  by the Hydrogen recombination, [\oi]6300\AA\, and some \feii\, lines. These profiles are characterized by a similar central and terminal velocity, which extends up to  $-$200 km s$^{-1}$.
\end{itemize}

\section{Acknowledgements}
This work is based on observations made with different instruments:[1] the LBT. The LBT is an international collaboration among institutions in the United States, Italy and Germany. LBT Corporation partners are: The University of Arizona on behalf of the Arizona university system; Istituto Nazionale di Astrofisica, Italy; LBT Beteiligungsgesellschaft, Germany, representing the Max-Planck Society, the Astrophysical Institute Potsdam, and Heidelberg University; The Ohio State University, and The Research Corporation, on behalf of The University of Notre Dame, University of Minnesota and University of Virginia; [2] the Asiago ANS Collaboration telescopes operated under the responsability of the INAF-Osservatorio Astronomico di Padova (OAPd); [3] the AZT-24 IR telescope at Campo Imperatore (L'Aquila - Italy) operated by INAF-Osservatorio Astronomico di Roma (OAR).
We acknowledge the observers who contribute to the AAVSO International Database. The authors are grateful to Lynn Hillenbrand for providing her spectra of V2492 Cyg.

\end{document}